\documentclass[12pt,preprint]{aastex}
\slugcomment{2011.5 v3}

\shorttitle{AXP/SGR gamma-ray emissions} \shortauthors{Tong, Song,& Xu}

\begin{document}

\title{AXPs and SGRs in the outer gap model: confronting Fermi observations}

\author{H. Tong\altaffilmark{1}
\email{haotong@ihep.ac.cn}}

\author{L. M. Song\altaffilmark{1} and R. X. Xu\altaffilmark{2}}

\altaffiltext{1}{Institute of High Energy Physics, Chinese Academy
of Sciences, Beijing, China}
\altaffiltext{2}{School of Physics and State Key
Laboratory of Nuclear Physics and Technology, Peking University, Beijing, China}

\begin{abstract}
Anomalous X-ray pulsars (AXPs) and soft gamma-ray repeaters (SGRs)
are magnetar candidates, i.e., neutron stars powered by strong
magnetic field. If they are indeed magnetars, they will emit
high-energy gamma-rays which are detectable by Fermi-LAT according to
the outer gap model. However, no significant detection is reported
in recent Fermi-LAT observations of all known AXPs and SGRs.
Considering the discrepancy between theory and observations, we
calculate the theoretical spectra for all AXPs and SGRs with
sufficient observational parameters. Our results show that most AXPs
and SGRs are high-energy gamma-ray emitters if they are really
magnetars. The four AXPs 1E 1547.0-5408, XTE J1810-197, 1E 1048.1-5937, and 4U
0142+61 should have been detected by Fermi-LAT. Then there is
conflict between out gap model in the case of magnetars and Fermi observations. 
Possible explanations in the magnetar model are discussed.
On the other hand, if AXPs and SGRs are fallback disk systems, i.e., accretion-powered
for the persistent emissions, most of them are not high-energy
gamma-ray emitters. Future deep Fermi-LAT observations of AXPs and
SGRs will help us make clear whether they are magnetars or fallback 
disk systems.
\end{abstract}


\section{Introduction}

Anomalous X-ray pulsars (AXPs) and soft gamma-ray repeaters (SGRs)
are two peculiar kinds of pulsar-like objects. Their persistent
X-ray luminosities are in excess of their rotational energy loss
rates, while at the same time they show no binary signature (review
Mereghetti 2008). They also show recurrent SGR-type bursts (review
Hurley 2009). Therefore, the energy budget of AXPs and SGRs is a
fundamental problem in their studies. They are supposed to be
magnetic field powered, i.e., magnetars (Thompson \& Duncan 1995,
1996). Another possibility is that they are accretion powered
systems, i.e., accretion from supernova fallback disks (Alpar 2001;
Chatterjee et al. 2000; Xu et al. 2006). Then, it is of fundamental
importance to determine whether they are magnetars or fallback disk systems. 
Solving this problem is also helpful to other high-energy astrophysical
phenomena and related pulsar-like objects (Xu 2007; Tong et al.
2010a)

Cheng \& Zhang (2001) proposed that although AXPs are slowly
rotating neutron stars, if their surface dipole magnetic field is
strong enough (i.e., if they are really magnetars) then they can
accelerate particles and emit high-energy gamma-rays which are
detectable by Fermi-LAT according to the outer gap model (Zhang \&
Cheng 1997). However, Sasmaz Mus \& Gogus (2010) reported a
non-detection in a Fermi-LAT observation of AXP 4U 0142+61. This
observation is in conflict with the outer gap
model. Tong et al. (2010b) proposed that Fermi-LAT observations can
help us distinguish between the magnetar model and the fallback disk 
model. Recently, the Fermi-LAT collaboration have published their
observations for all known AXPs and SGRs (five SGRs and eight AXPs),
where still no significant detection is reported (Abdo et al.
2010b). Considering this discrepancy between theory and
observations, it is then very necessary to do a comprehensive study
of this issue.

In Cheng \& Zhang (2001), only five AXPs are considered and the
paremeters they used are very uncertain, e.g., the surface
temperatures are estimated from the X-ray luminosities, etc. Now, we
have very good observational data for more sources (see the McGill
AXP/SGR online catalog). On the other hand, there are also
developments of the outer gap model (e.g., Takata et al. 2010). In
this paper, with up-to-date observational parameters of AXPs and
SGRs, we consider the high-energy gamma-ray radiation properties of
AXPs and SGRs in the outer gap model (Zhang \& Cheng 1997;
Takata et al. 2010) and compare them with Fermi-LAT observations.

Section 2 is application of self-consistent outer gaps to AXPs and
SGRs. We consider both the magnetar model and fallback disk model.
Discussions and conclusions are presented in Section 3 and Section
4, respectively.

\section{Application of self-consistent outer gaps to AXPs and SGRs}

The outer gap is very successful in explaining pulsar high-energy emissions (Cheng et al. 1986; review Cheng 2009).
Zhang \& Cheng (1997) developed the self-consistent outer gap model where the longitudinal extension
of outer gap is determined self-consistently by the $\gamma-\gamma$ pair production process. If the X-ray photons
are provided by neutron star surface thermal emission, the size of outer gap is (Zhang \& Cheng 1997)
\begin{equation}
 f_{\gamma\gamma}=4.5 P^{7/6} B_{12}^{-1/2} T_{6}^{-2/3} R_{6}^{-3/2},
\end{equation}
where $P$ is the neutron star rotation period, $B_{12}$ is the surface magnetic field in units of $10^{12}\,\mathrm{G}$,
$T_{6}$ is the surface temperature in units of $10^6\,\mathrm{K}$, and $R_{6}$ is the neutron star radius in units of
$10^{6}\,\mathrm{cm}$. Here, $f$ should be less than one for outer gap to exist. Takata et al. (2010)
further considered the $\gamma-$B pair production process as a gap closure mechanism. The size of outer gap at half the light cylinder radius
is (Takata et al. 2010)
\begin{equation}\label{fm}
f_{\mathrm{m}}= 2^{-3/2}\times 0.25 K(\chi, B_{\mathrm{m}},s) P_{-1}^{1/2},
\end{equation}
where $P_{-1}$ is the rotation period in units of 0.1 seconds, $K$
depends on the local geometry of magnetic fields at which the $\gamma-$B process takes place (Takata et al. 2010).
\begin{equation}
 K= \chi_{-1}^2 B_{\mathrm{m},12}^{-2} s_7 (R/R_{\mathrm{i}})^{3/2},
\end{equation}
where $\chi_{-1}$ is a dimensionless parameter in units of $0.1$, which depends on the angle between photon propagation
direction and magnetic field, $B_{\mathrm{m},12}$
is the multipole field in units of $10^{12}\,\mathrm{G}$, $s_7$ is
the local curvature radius in units $10^{7}\,\mathrm{cm}$, $R$ is
the neutron star radius, and $R_{\mathrm{i}}$ is the radial distance
at which the $\gamma-$B process takes place.

\subsection{Calculations in the case of magnetars}

With up-to-date observational parameters of AXPs and SGRs,
we have calculated the gamma-ray radiation properties of all AXPs and SGRs which have period, period derivative,
surface temperature, and distance measurement (except one source in SMC).
Three SGRs and ten AXPs (including two candidates) are selected.
The results are summarized in Table 1.
The period, period derivative, surface temperature, and distance data are all
from the McGill AXP/SGR catalog\footnote{http://www.physics.mcgill.ca/$\sim$pulsar/magnetar/main.html, up to February 9, 2011}
(except the distance data of SGR 0501+4516 which is from Abdo et al. (2010b) and reference therein).
The magnetic field is calculated from $B=6.4\times10^{19}\sqrt{P\dot{P}}$, which is 2 times larger than usually reported since
polar magnetic field is more important in the case of pulsar radiation (Shapiro \& Teukolsky 1983).
We consider the typical case with inclination angle equals $60^{\circ}$ (Cheng \& Zhang 2001).
The solid angle is chosen as $\Delta \Omega=1$.
A star radius $R=12\,\mathrm{km}$ is employed, which corresponds to medium to stiff equation of state.
A medium to stiff equation of state is favored by the recently measured 2 solar-mass neutron star (Demorest et al. 2010).

For a magnetar, whose surface magnetic field is about $10^{15}\,\mathrm{G}$, the $\gamma-$B pair production process
will take place at about 10 stellar radius where the magnetic field is about $10^{12}\,\mathrm{G}$.
Then the K-parameter in Equation (\ref{fm}) is about 2. Since magnetars are slowly rotating neutron stars,
the size of outer gap $f_{\mathrm{m}}$ will always be larger than one. Therefore, for magnetars, the gap closure
mechanism will be dominated by the $\gamma-\gamma$ pair production process. This conclusion is depicted quantitively
in Table 1. From Table 1, we see that only for one AXP 1E 2259+586, the size of outer gap is larger than one.
Therefore, this AXP will not emit high-energy gamma-rays. For the rest of AXPs and SGRs, they are all
high-energy gamma-ray emitters according to the outer gap model (Zhang \& Cheng 1997).

The theoretical spectra energy distributions (SEDs) are calculated following Zhang \& Cheng (1997), Cheng \& Zhang (2001).
For the three SGRs and nine AXPs (including two candidate AXPs), their spectra are shown in Figure 1 and 2, and summarized
in Table 1. We see that due to their large distances, the three SGRs (SGR 1806-20, SGR 1900+14, and SGR 0501+4516)
and four AXPs (CXO J164710.2-455216, 1RXS J170849.0-400910, 1E 1841-045, and PSR J1622-4950) can not be detected
by Fermi-LAT for one year exposure time, i.e., their SEDs lie below the Fermi-LAT sensitivity
curve. For CXOU J171405.7-381031, its SEDs lie in the vicinity of Fermi-LAT sensitivity curve.
Therefore, the detectability is only marginal.
The most notable exceptions are 1E 1547.0-5408, XTE J1810-197, 1E 1048.1-5937, and 4U 0142+61, whose SEDs lie well above the Fermi-LAT
sensitivity curve. Therefore, they should be detected by Fermi-LAT observations.

\begin{table}[t]
\footnotesize
\centering
\caption{\footnotesize Size of outer gap for three SGRs and ten AXPs.
Column one to eight are source name, period, period derivative, surface temperature, distance,
size of outer gap $f_{\gamma\gamma}$, size of outer gap $f_{\mathrm{m}}$, and its detectability by
Fermi-LAT for one year exposure time. The two candidate AXPs PSR J1622-4950, and CXOU J171405.7-381031 are also included.
All data are from the McGill AXP/SGR catalog
(except the distance data of SGR 0501+4516 which is from Abdo et al. (2010b) and reference therein).
}

\begin{tabular}{llllllll}
\tableline\tableline
Source & P & $\dot{P}$ & $T_{\mathrm{BB}}$ & d  & $f_{\gamma\gamma}^c$ &  $f_{\mathrm{m}}$ & Detectability\\
 & (sec) & ($10^{-11}$) & (keV) & (kpc) & & & \\
\tableline
SGR 1806-20 & 7.6022 & 75 & 0.6 & 8.7 & 0.14 (0.19) & 1.54 & NO\\

SGR 1900+14 & 5.1999 & 9.2 & 0.47 & 13.5$^a$ & 0.20 (0.27) &  1.27 & NO\\

SGR 0501+4516 & 5.7621 & 0.582 & 0.69 & 5.0 & 0.34 (0.45) & 1.34 & NO\\

1E 1547.0-5408 & 2.0698 & 2.318 & 0.43 & 3.9 & 0.13 (0.17) & 0.80 & YES\\

XTE J1810-197 & 5.5404 & 0.777 & $0.301^{d}$ & 3.5 & 0.54 (0.70) & 1.32 & YES\\

1E 1048.1-5937 & 6.4521 & 2.70 & 0.623 & 2.7 & 0.28 (0.37) & 1.42 & YES\\

1E 2259+586 & 6.9789 & 0.048 & 0.411 & 4.0 & 1.1 (1.4) & 1.48 & Never\\

4U 0142+61 & 8.6883 & 0.196 & 0.395 & 2.5$^b$ & 0.95 (1.3) & 1.65 & YES\\

CXO J164710.2-455216 & 10.6107 & 0.24 & 0.63 & 5 & 0.80 (1.05) & 1.82 & NO\\

1RXS J170849.0-400910 & 10.999 & 1.945 & 0.456 & 8 & 0.61 (0.80) & 1.85 & NO\\

1E 1841-045 & 11.775 & 4.1551 & 0.44 & 8.5 & 0.55 (0.72) & 1.92 & NO\\

PSR J1622-4950 & 4.3261 & 1.7 & 0.4 & 9 & 0.29 (0.38) & 1.16 & NO\\

CXOU J171405.7-381031 & 3.8254 & 6.40 & 0.38 & 8 & 0.19 (0.25) & 1.09 & Marginal \\

\tableline
\end{tabular}
\flushleft
\textbf{Notes:} %

$^a$: median value is employed

$^b$: lower limit is employed

$^c$: $f_{\gamma\gamma}$ when star radius $R=12\,\mathrm{km}$ ($R=10\,\mathrm{km}$ in brackets)

$^d$: the temperature of the hotter component is employed

\end{table}

\begin{figure}[t]
\centering
\begin{minipage}{0.45\textwidth}
\includegraphics[height=0.3\textheight]{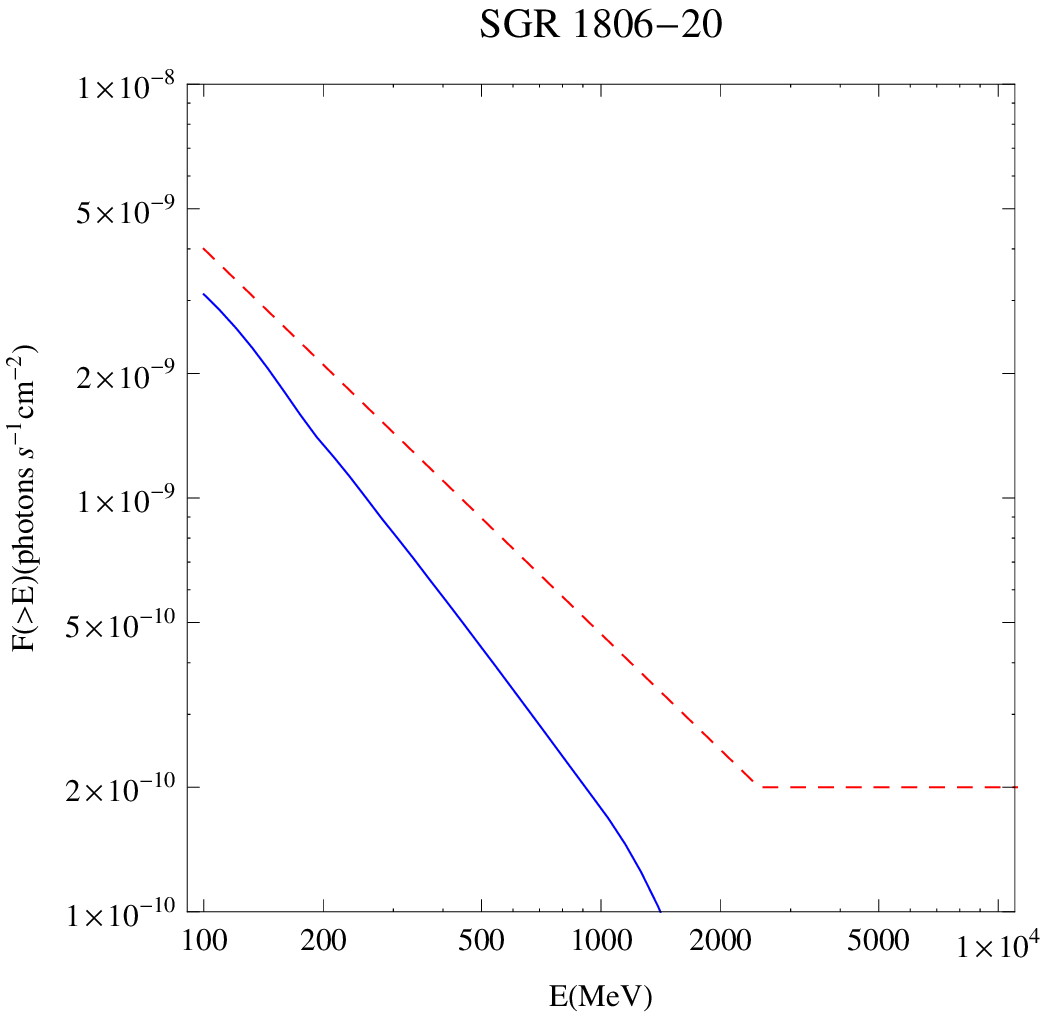}
\end{minipage}
\begin{minipage}{0.45\textwidth}
\includegraphics[height=0.3\textheight]{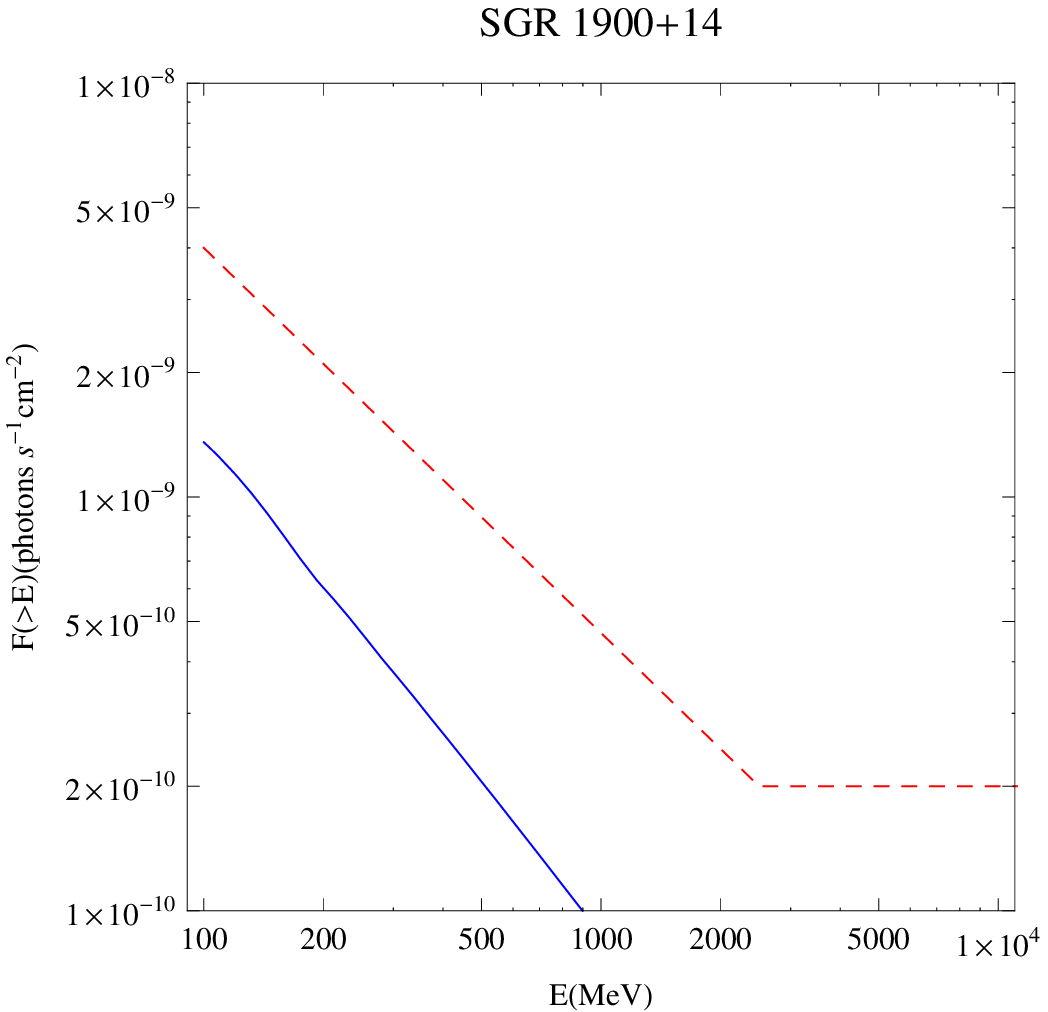}
\end{minipage}

\begin{minipage}{0.45\textwidth}
\includegraphics[height=0.3\textheight]{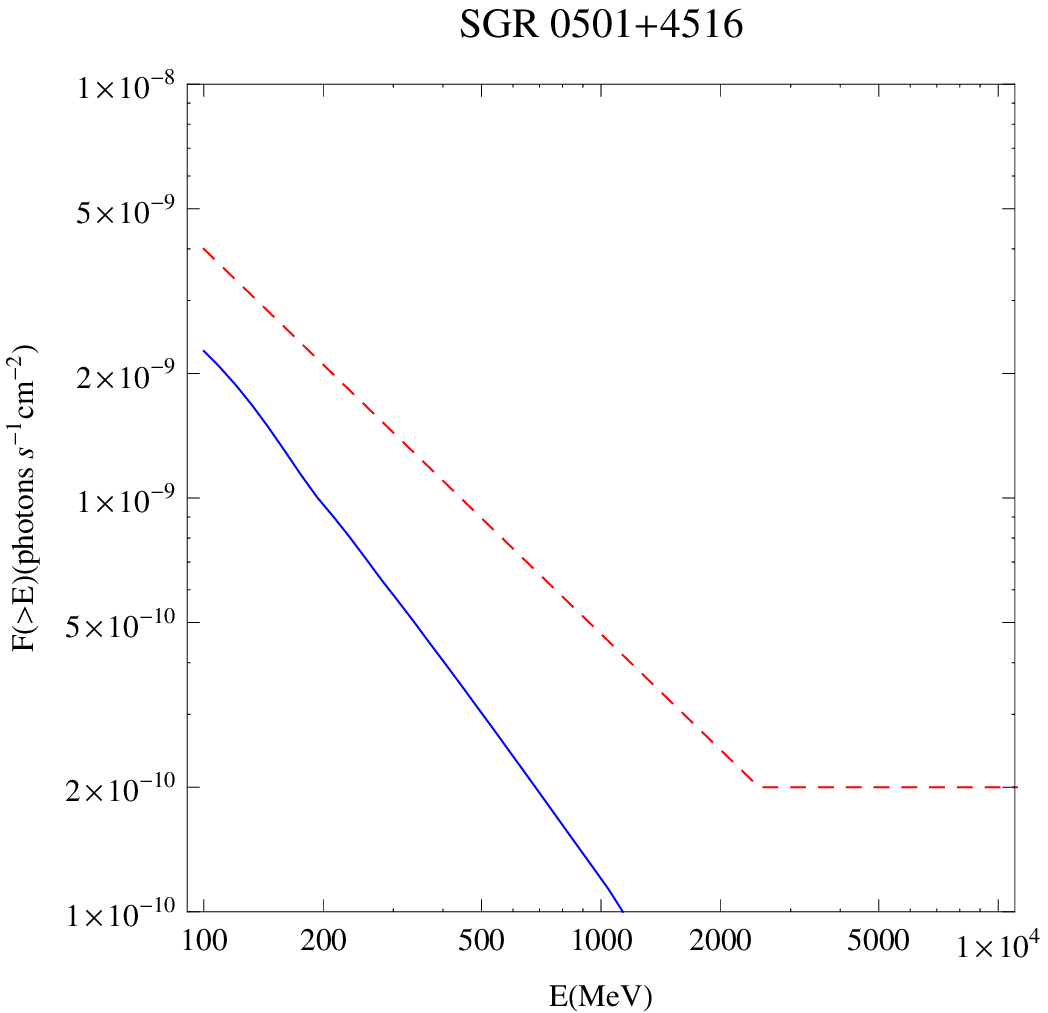}
\end{minipage}
\begin{minipage}{0.45\textwidth}
\includegraphics[height=0.3\textheight]{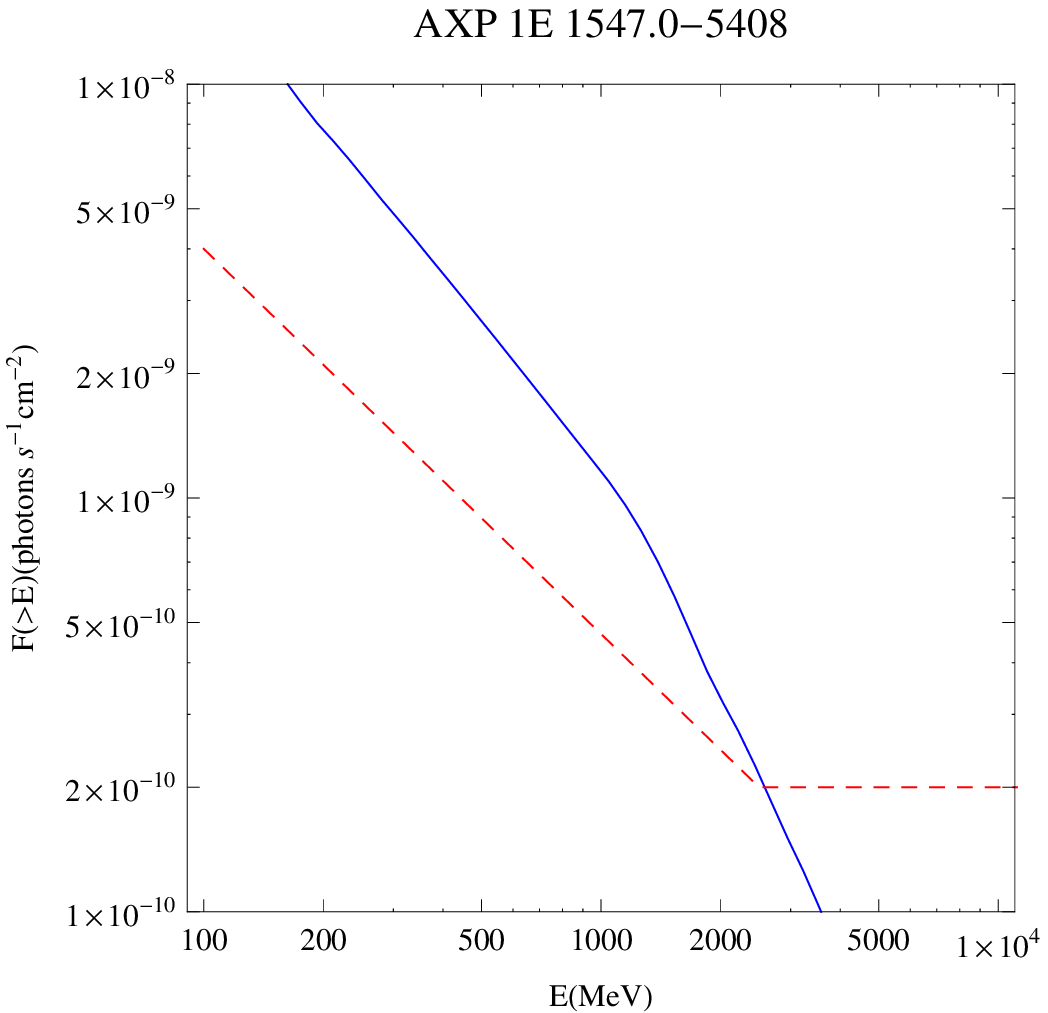}
\end{minipage}

\begin{minipage}{0.45\textwidth}
\includegraphics[height=0.3\textheight]{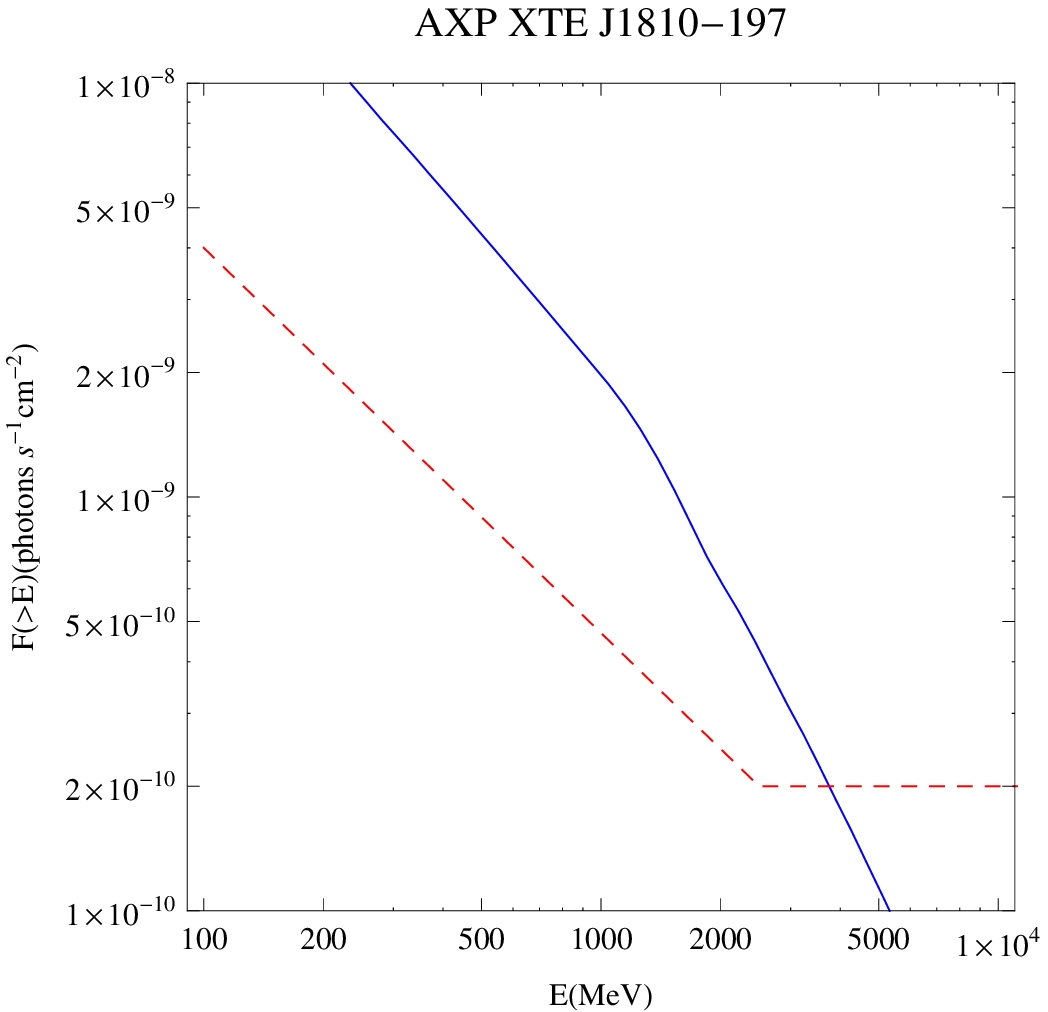}
\end{minipage}
\begin{minipage}{0.45\textwidth}
\includegraphics[height=0.3\textheight]{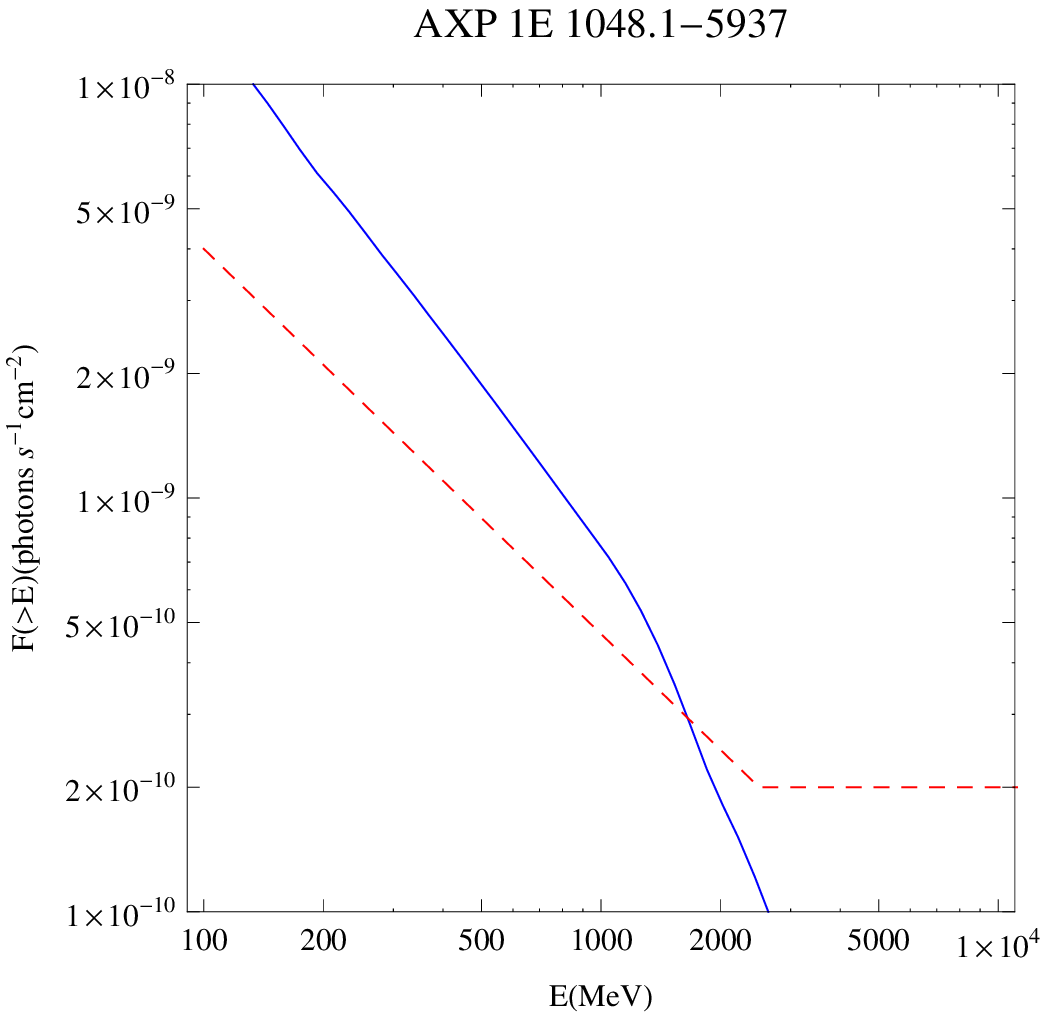}
\end{minipage}

\caption{Integral spectra vs. Fermi-LAT sensitivity curve. The solid
line is the theoretical spectra according to the out gap model
(Zhang \& Cheng 1997; Cheng \& Zhang 2001). The dashed line is the Fermi-LAT sensitivity
curve for one year exposure time (Atwood et al 2009). Typical
calculations for SGR 1806-20, SGR 1900+14, SGR 0501+4516, 1E
1547.0-5408, XTE J1810-197, and 1E 1048.1-5937.}
\end{figure}

\begin{figure}[t]
\centering
\begin{minipage}{0.45\textwidth}
\includegraphics[height=0.3\textheight]{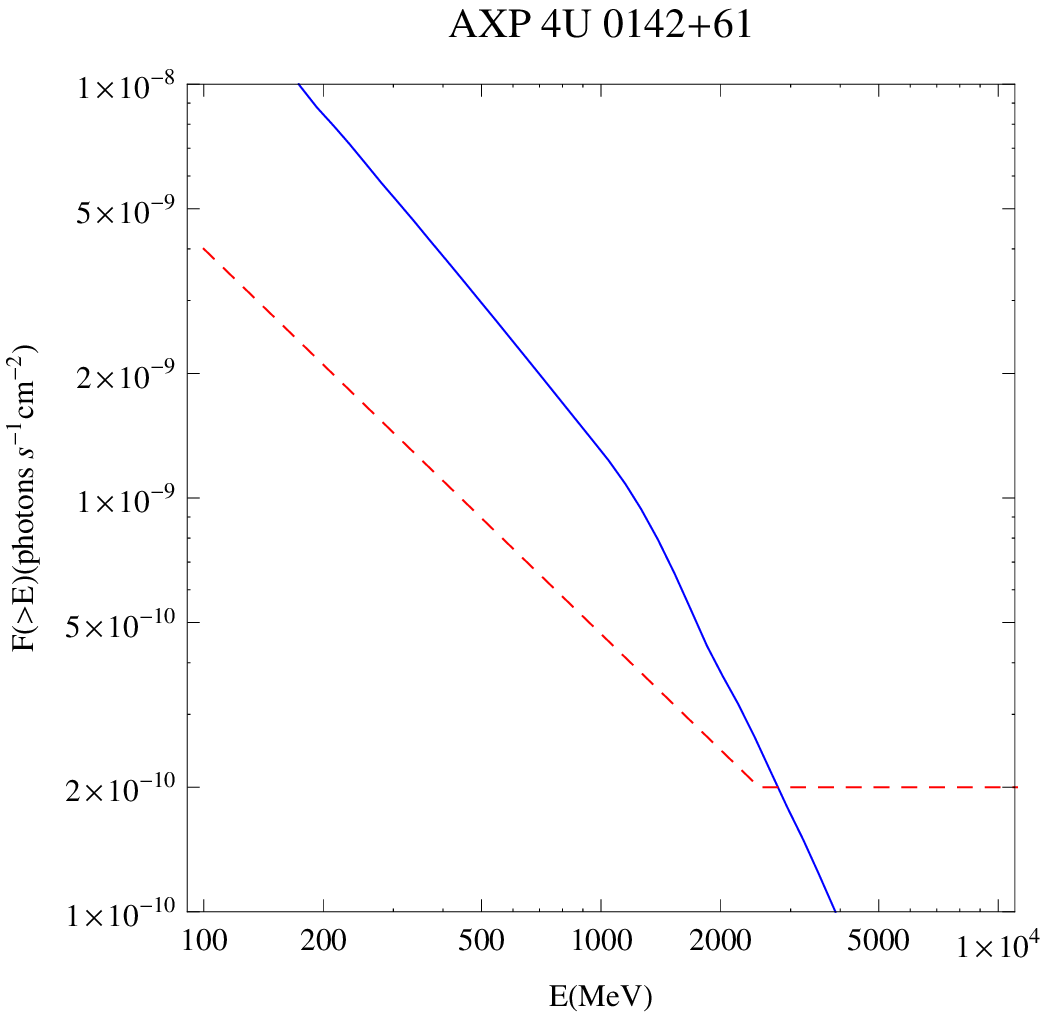}
\end{minipage}
\begin{minipage}{0.45\textwidth}
\includegraphics[height=0.3\textheight]{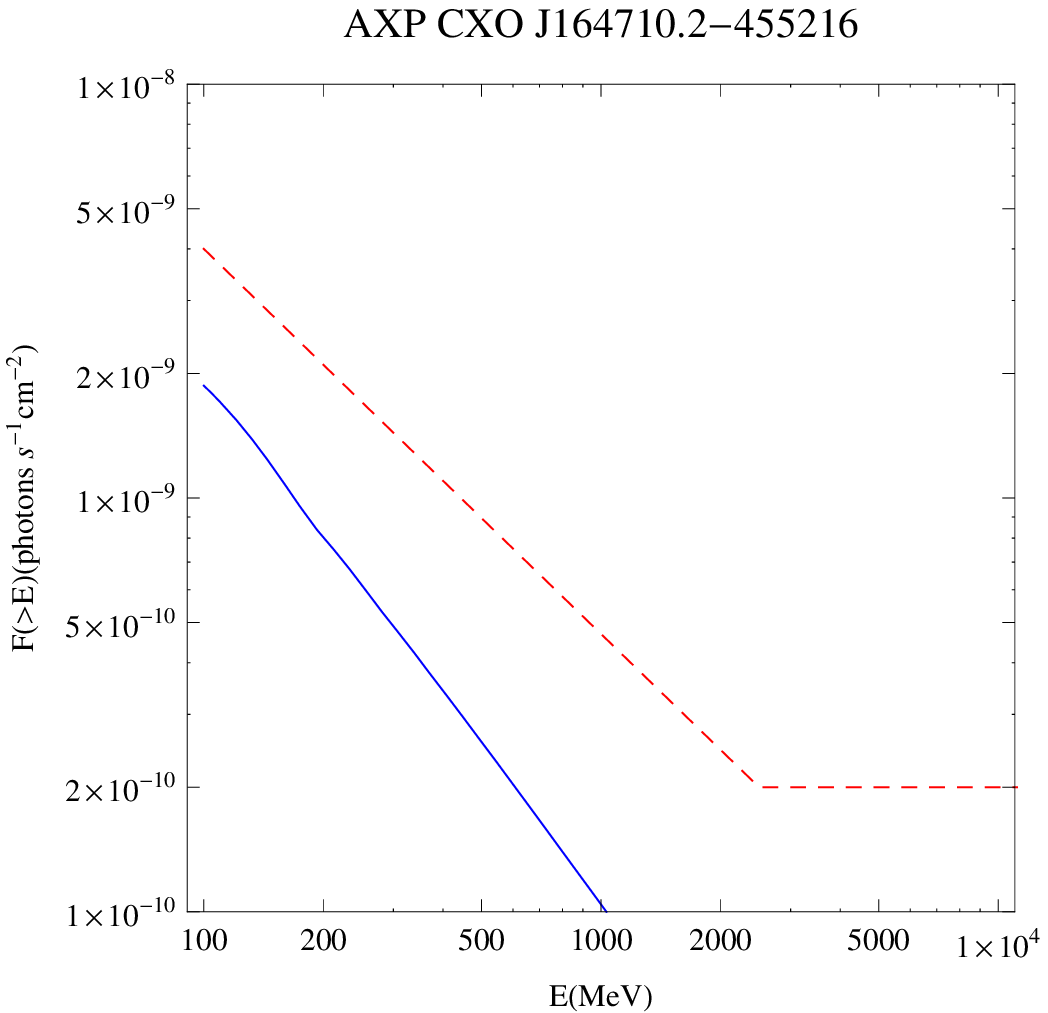}
\end{minipage}

\begin{minipage}{0.45\textwidth}
\includegraphics[height=0.3\textheight]{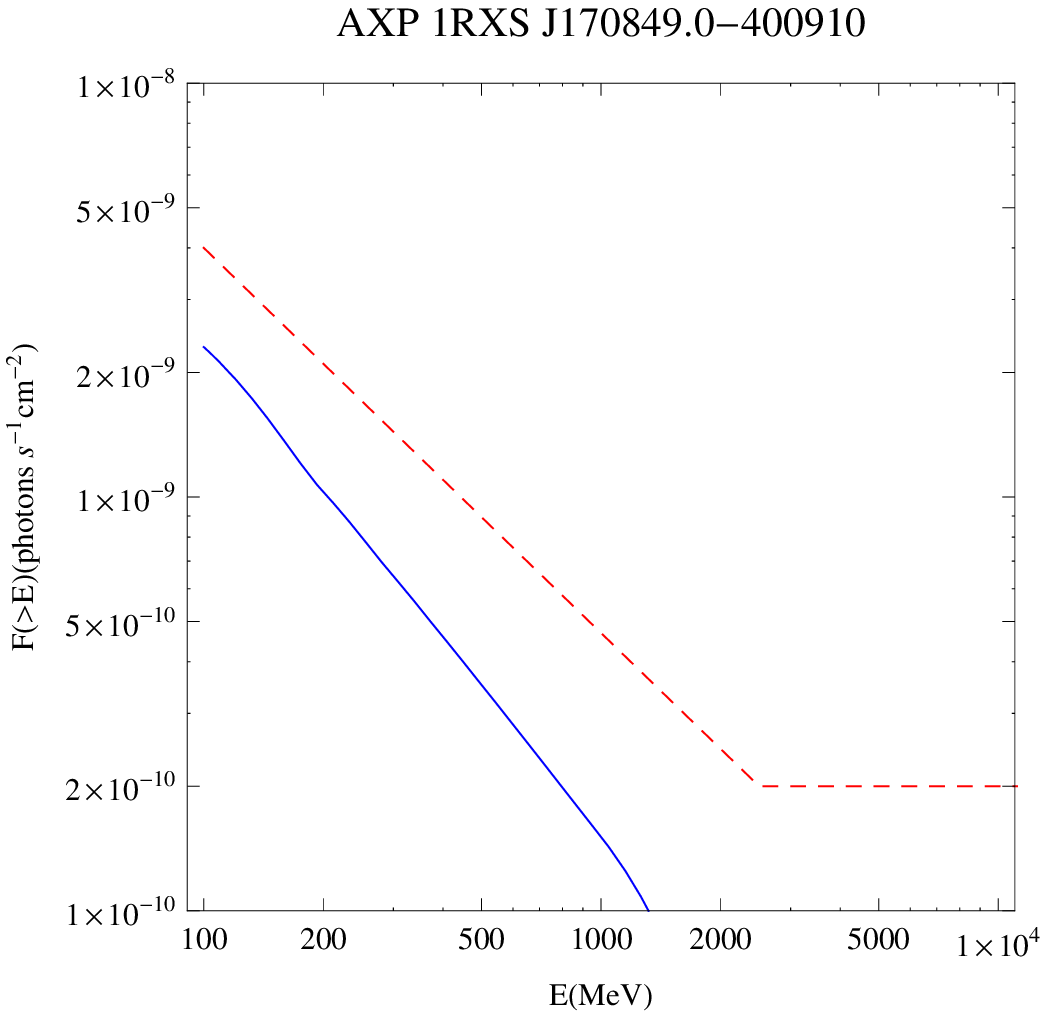}
\end{minipage}
\begin{minipage}{0.45\textwidth}
\includegraphics[height=0.3\textheight]{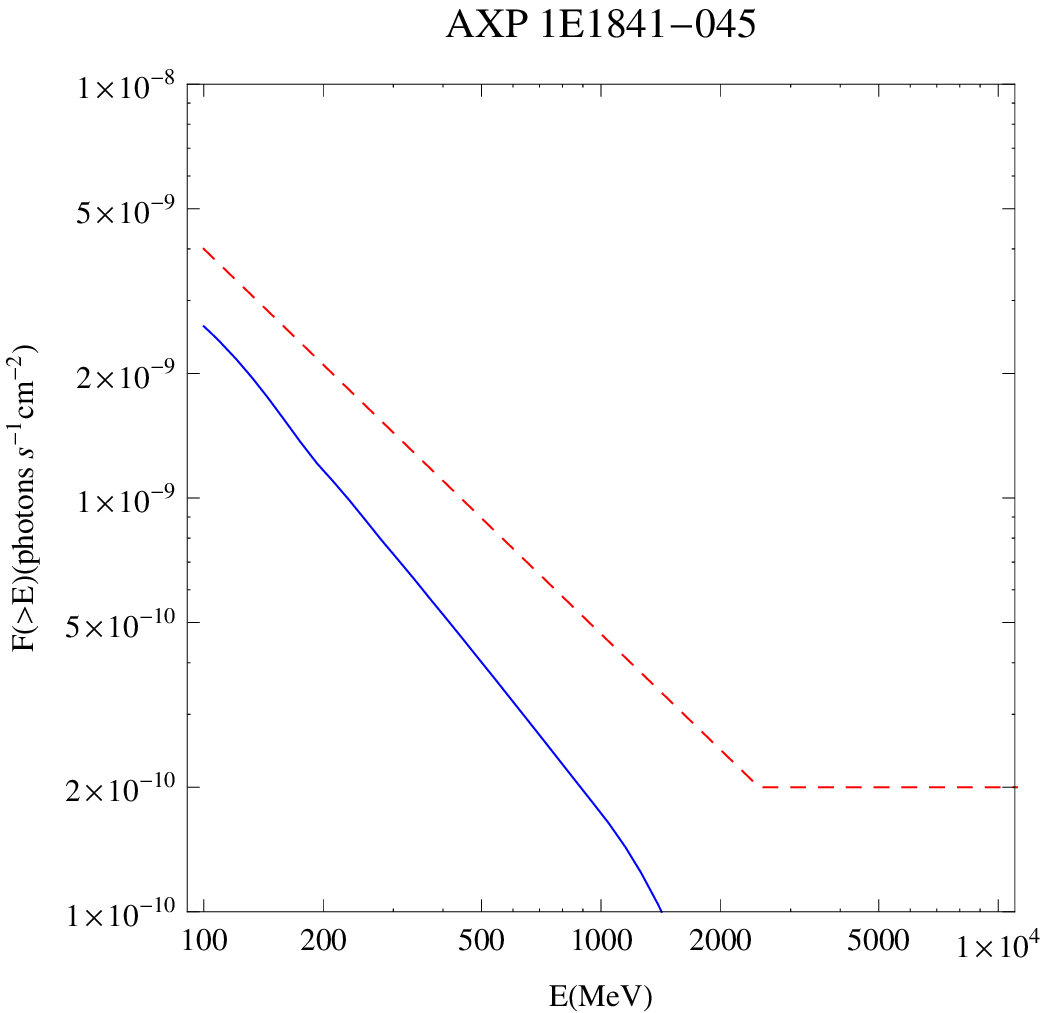}
\end{minipage}

\begin{minipage}{0.45\textwidth}
\includegraphics[height=0.3\textheight]{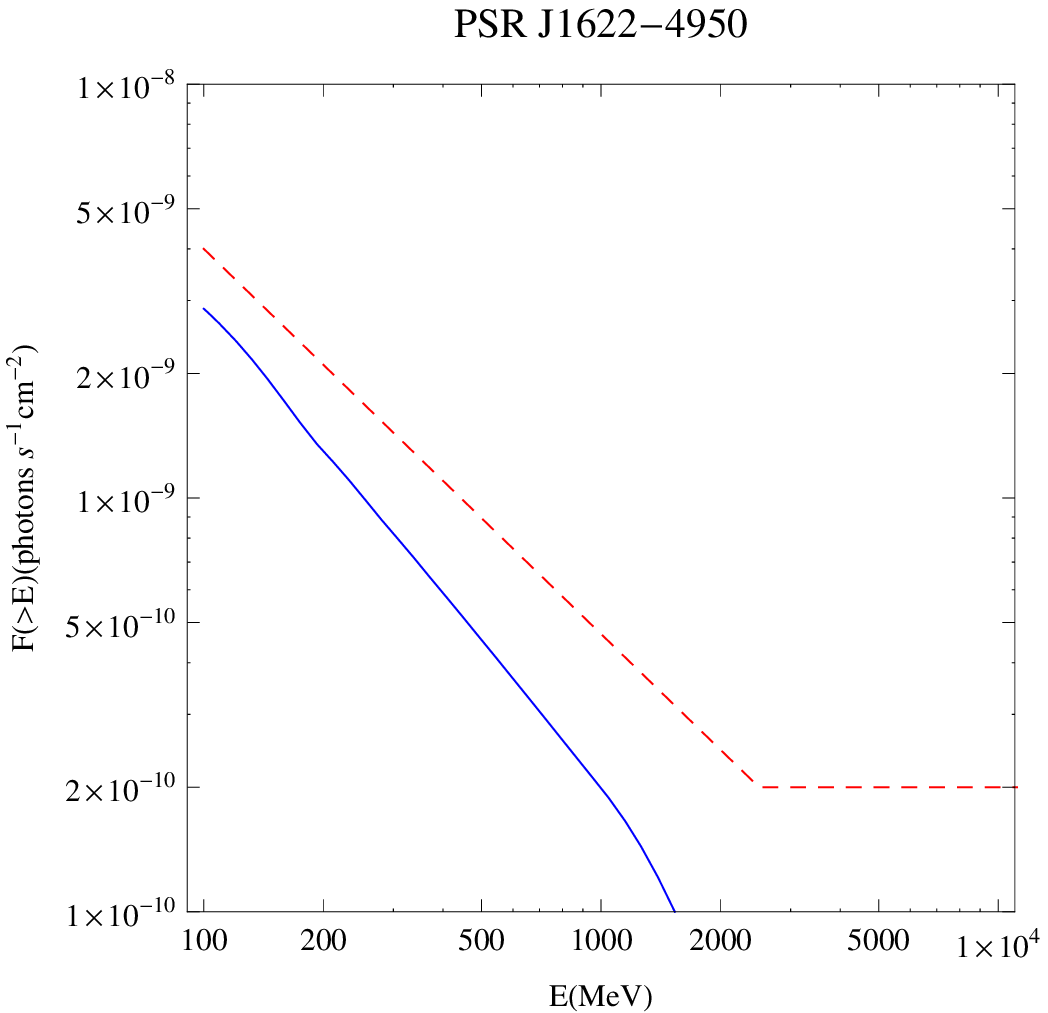}
\end{minipage}
\begin{minipage}{0.45\textwidth}
\includegraphics[height=0.3\textheight]{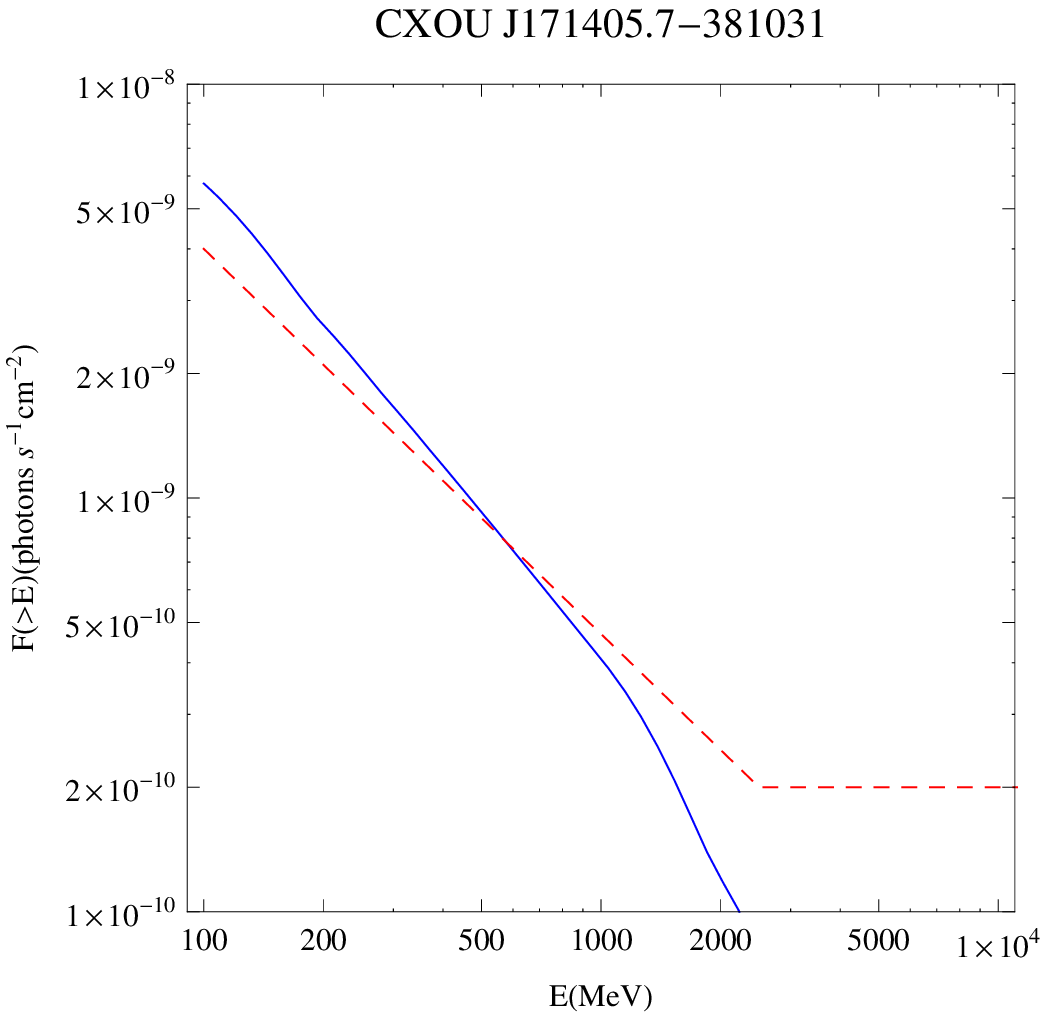}
\end{minipage}

\caption{Integral spectra vs. Fermi-LAT sensitivity curve. Typical calculations
for 4U 0142+61, CXO J164710.2-455216, 1RXS J170849.0-400910, 1E 1841-045, PSR J1622-4950, and
CXOU J171405.7-381031.}
\end{figure}

\subsection{Comparison with Fermi-LAT observations}

The Fermi-LAT collaboration have published their observations for all known AXPs and SGRs
(Abdo et al. 2010b, five SGRs and eight AXPs). Despite 17 months of Fermi-LAT
observations, no significant detection is reported. For the three SGRs and eight AXPs
considered in this paper in Table 1 (except two candidate AXPs), they are all observed by Fermi-LAT
(cf. Table 1 in Abdo et al. 2010b).

For 1E 2259+586, it will not emit high-energy gamma-rays according to the outer gap model (Zhang \& Cheng 1997).
For the three SGRs (SGR 1806-20, SGR 1900+14, and SGR 0501+4516)
and three AXPs (CXO J164710.2-455216, 1RXS J170849.0-400910, and 1E 1841-045), mainly due to their large distances,
they can not be detected by Fermi-LAT for one year exposure time (17 months exposure time will not make qualitative improvements).
Therefore, for these seven sources, current Fermi-LAT observations can not put constraints
on theoretical models, i.e., they can be either magnetars or fallback disk systems (see Section 2.4).

Notable exceptions are 1E 1547.0-5408, XTE J1810-197, 1E 1048.1-5937, and 4U 0142+61, 
which should have been detected by Fermi-LAT for 17 months observations. 
Therefore, there are conflicts between outer gap predictions in the case of magnetars and
Fermi-LAT observations. As noticed by Tong et al. (2010b) for the single case of 4U 0142+61,
the non-detection in Fermi-LAT observations may propose challenges to the magnetar model. Now, the conflicts between theory
and observations are more severe for these four AXPs. We also compare the theoretical SEDs of these four sources with
their observational upper limits, shown in Figure 3. The upper limits for 1E 1547.0-5408, XTE J1810-197, and 1E 1048.1-5937 can not
provide strong constraints at present. The upper limits for 4U 0142+61 lie already below the theoretical SEDs for large
inclination angles.

\begin{figure}[t]
\centering
\begin{minipage}{0.45\textwidth}
\includegraphics[height=0.3\textheight]{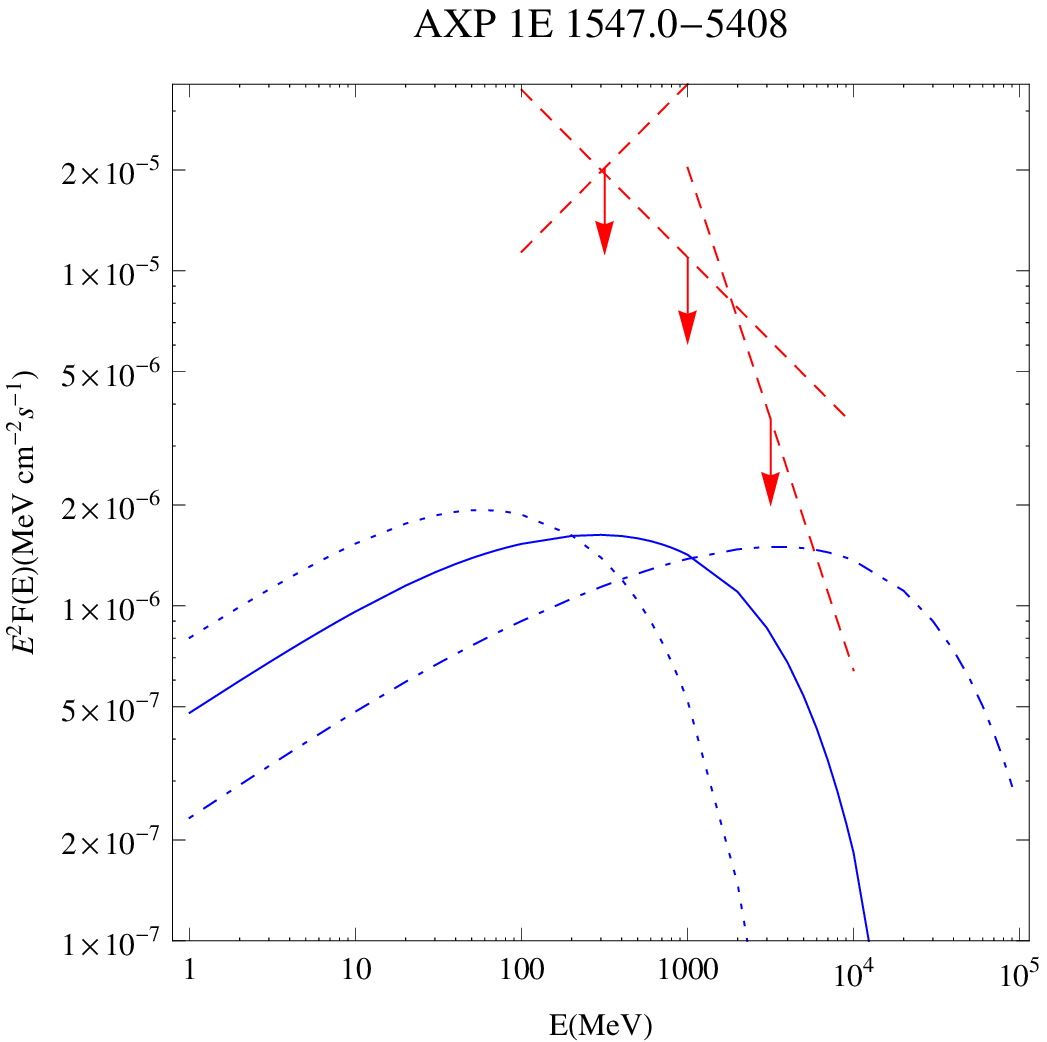}
\end{minipage}
\begin{minipage}{0.45\textwidth}
\includegraphics[height=0.3\textheight]{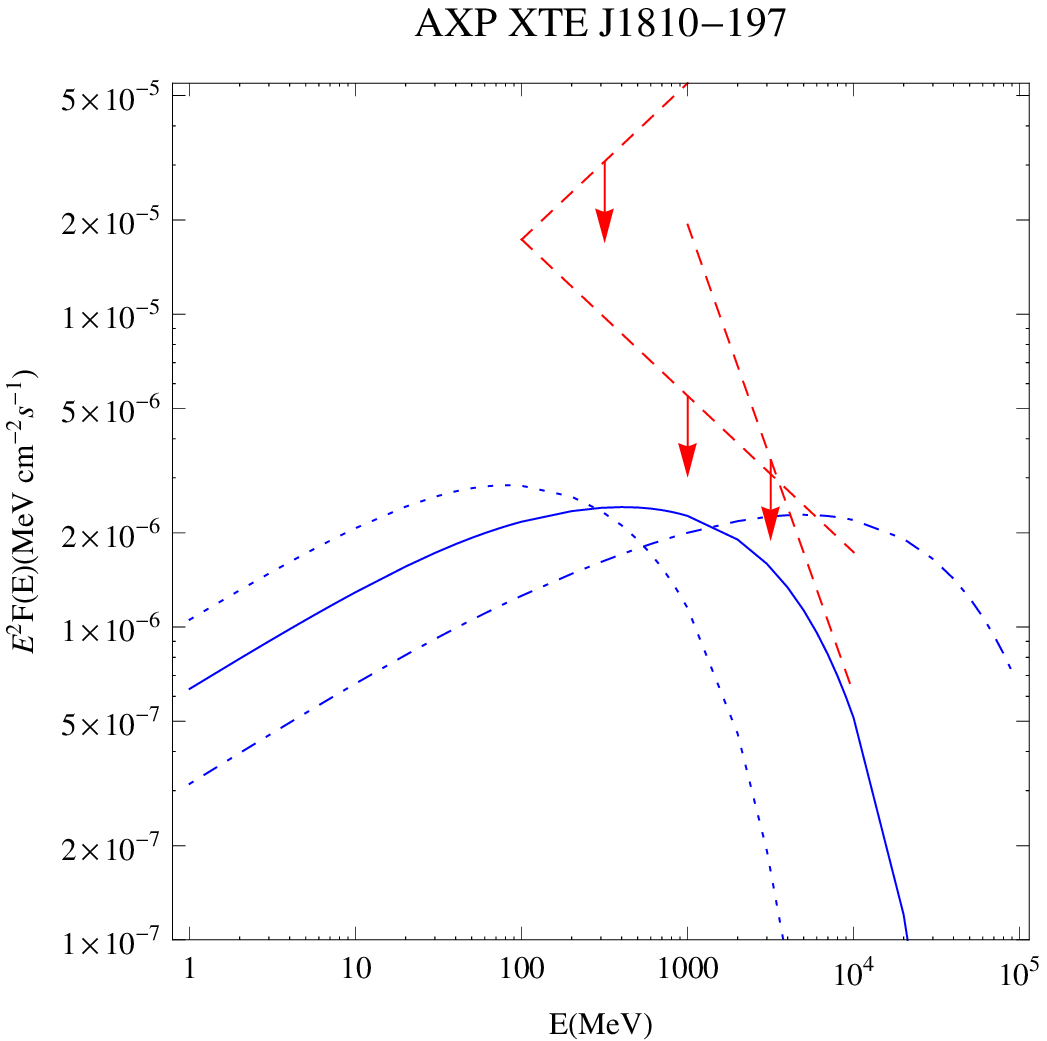}
\end{minipage}

\begin{minipage}{0.45\textwidth}
\includegraphics[height=0.3\textheight]{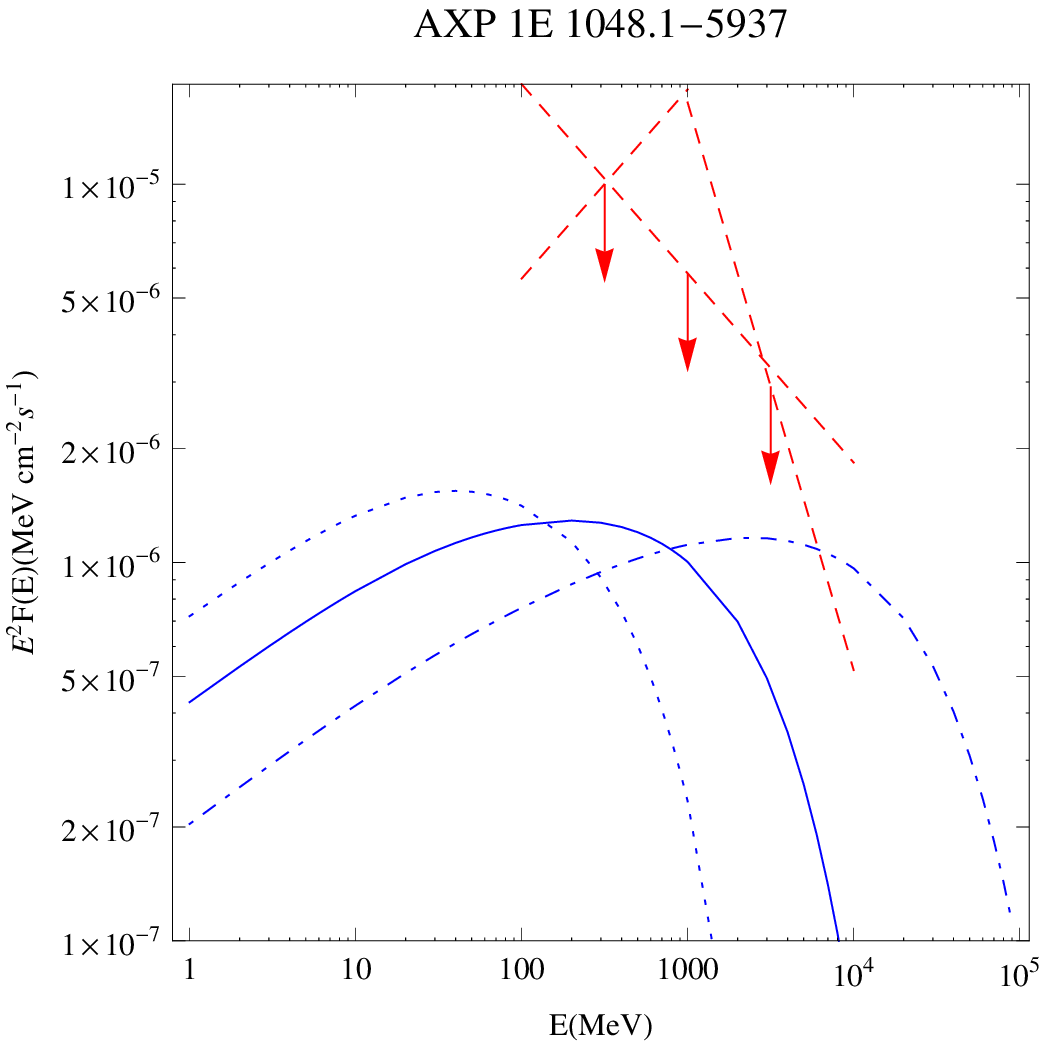}
\end{minipage}
\begin{minipage}{0.45\textwidth}
\includegraphics[height=0.3\textheight]{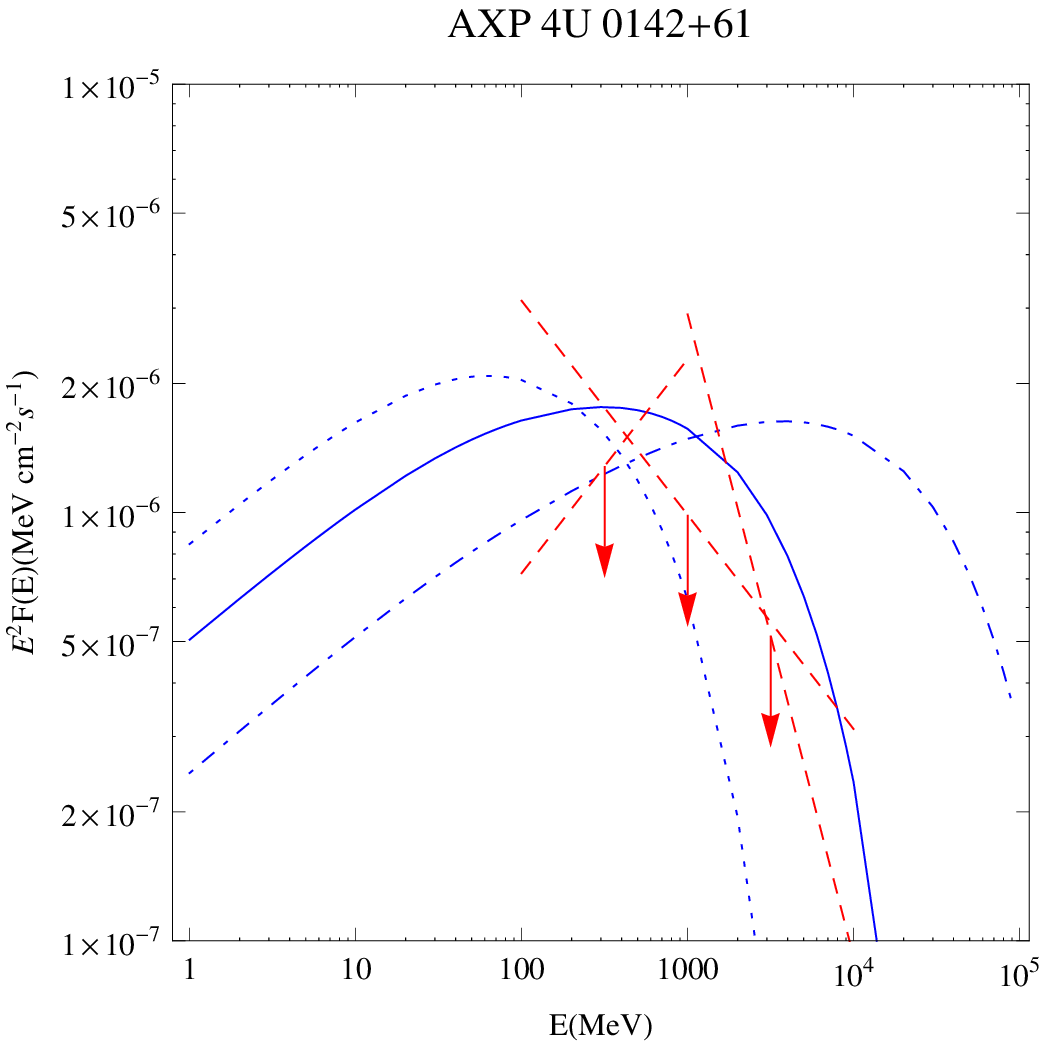}
\end{minipage}

\caption{Theoretical spectra energy distributions vs. Fermi-LAT
upper limits. The dotted, solid, and dot-dashed lines are
theoretical spectra for inclination angle $45^{\circ}$,
$60^{\circ}$, $75^{\circ}$ respectively (Zhang \& Cheng 1997; Cheng
\& Zhang 2001). The dashed lines are Fermi-LAT upper limits in
energy ranges 0.1-10 GeV, 0.1-1 GeV, and 1-10 GeV (Abdo et al.
2010b). We only show the calculations for 1E 1547.0-5408, XTE J1810-197, 
1E 1048.1-5937, and 4U 0142+61, which should have been detected by Fermi-LAT 
for one year exposure time.}
\end{figure}

\subsection{The applicability of outer gap model to magnetars}

In the magnetar model for AXPs and SGRs, both the bursts and
persistent emissions are powered by magnetic field (Thompson \&
Duncan 1995, 1996). While, the self-consistent outer gap model is
original designed for rotation powered pulsars (Zhang \& Cheng
1997). Therefore, the applicability of outer gap model to magnetars
may seem not very convincing at first sight. However, when deducing
the star magnetic field from timing observations, the magnetic
dipole braking mechanism is employed as in the case of rotation
powered pulsars (e.g., Kouveliotou et al. 1998). The consequence of
magnetic dipole braking is that the rotational
energy of AXPs and SGRs are carried away by similar processes to
that of rotation powered pulsars. Therefore, there should be some
rotation powered activities in magnetars and the high-energy
gamma-ray emissions are just one of them (Zhang 2003). The
high-energy gamma-ray properties of AXPs and SGRs discussed in
previous sections are the consequences of their strong surface
dipole field.

The magnetosphere of magnetars may be more complicated than
that of rotation powered pulsars, e.g., it may be twisted (Thompson
et al. 2002). A twisted magnetosphere contains higher multipoles in
addition to a dipole component. Far away in the outer magnetosphere
(as in the case of outer gap model), the higher multipoles will be
suppressed dramatically. Thus the dipole component will dominate in
the outer magnetosphere of magnetars. The magnetic field strength
there is below the quantum critical value. Therefore, we can apply
the outer gap model to magnetars (Cheng \& Zhang 2001). Considering
the detailed electrodynamics of magnetars, the magnetic field is
only quantitatively stronger than the dipole case (Thompson et al.
2002). The corotation charge density now has an
extra term in addition to the Goldreich-Julian term $\rho=
\rho_{\mathrm{GJ}} +\rho_{\mathrm{twist}}$ (Thompson et al. 2002). However, the twist term is only present in
closed field line regions in the vicinity of neutron star surface
where the magnetic field is strong and highly twisted (Beloborodov \& Thompson 2007). In the outer
magnetosphere where the magnetic field has been decreased greatly,
the Goldreich-Julian term will dominate and we expect there will
also be null charge surfaces as in the case of rotation powered
pulsars. The existence of null charge surface ensures the existence
of outer gaps (Cheng et al. 1986). 

Simulations of pair cascades
in strong magnetic field show that the main results are not strongly affected by 
photon splitting and a magnetar strength field but only dependent on 
the acceleration potential (Medin \& Lai 2010). 
The the maximum acceleration potential from a
rotating dipole is $6.6\times 10^{12}B_{12}P^{-2}\,\mathrm{V}$
(Cheng 2009). For AXPs and SGRs whose rotation period is about 10
seconds, if their surface magnetic fields are about
$10^{15}\,\mathrm{G}$, then they may accelerate particles to high
enough energy and emit high-energy gamma-rays.

In conclusion, the applicability of outer gap model to
magnetars is plausible. AXPs and SGRs are high-energy gamma-ray
emitters in the magnetar model because they have strong surface
dipole field.

\subsection{The case of fallback disk systems}

We now consider the possibility that AXPs and SGRs are fallback disk
systems (Alpar 2001; Chatterjee et al. 2000). The accretion
flow will quench the magnetospheric activities of the putative
neutron star. The radiation due to accretion will be mainly in the
soft X-ray and hard X-ray band (Frank et al. 2003). Recent fallback
disk modeling of AXPs and SGRs can explain their soft and hard X-ray
spectra uniformly (Trumper et al. 2010). Fermi-LAT observations of
the most luminous AXP 4U 0142+61 also indicate an energy break at
about $1\, \mathrm{MeV}$(Sasmaz Mus \& Gogus 2010; Abdo et al.
2010b).

High-energy gamma-ray radiation of AXPs and SGRs in the
fallback disk case are considered by Ertan \& Cheng (2004) in the disk-star dynamo
model (they only calculate the case of AXP 4U 0142+61). 
In order to generate high-energy gamma-rays, the inner disk
has to rotate faster than the neutron star. However, this can not be
fulfilled for the debris disk around AXP 4U 0142+61 either as a
passive disk (Wang et al. 2006) or a gaseous accretion disk (Ertan et al. 2007). On the
other hand, the outer gap is not supposed to operate in the fallback
disk case mainly due to the dense accretion flow. Therefore, AXPs
and SGRs are not high-energy gamma-ray emitters if they are fallback
disk systems.

\section{Discussions}

Our calculations show that the gap closure mechanism is dominated by $\gamma-\gamma$ pair
production process in the case of magnetars. The seed X-ray photons are provided by surface thermal
emission. Observationally there is also a power law component of soft X-ray photons (Mereghetti 2008).
The inclusion of power law soft X-ray photons will enhance the magnetospheric activities
of magnetars (Zhang \& Cheng 2002). However, physical modeling of the power law component shows that
it is also of thermal origin both in the case of magnetars (Lyutikov \& Gavriil 2006; Tong et al. 2010a)
and in the case of fallback disk systems (Trumper et al. 2010).

The inclination angle is the main factor determining
the spectra shape for different sources, see Figure 3. The larger the inclination angle,
the harder the gamma-ray spectra. Modern outer gap model shows that
the outer gap may extend below the null charge surface (Hirotani 2006). Tong et al. (2010b) try to
take this effect into consideration when calculating the gamma-ray spectra.
The corresponding spectra is similar to the case of large inclination angles,
e.g., $75^{\circ}$.

During our calculations, the solid angle $\Delta\Omega$ is chosen as unity
which is usually assumed (Cheng \& Zhang 2001 and references therein).
Outer gap modeling of Fermi gamma-ray pulsars also gives an average solid angle
of order unity (Wang et al. 2010).

The AXPs and SGRs lie mainly in the Galatic plane. The Fermi-LAT
threshold sensitivity may be 3-5 times larger in the Galatic plane
than that at higher latitude as in the case of gamma-ray pulsars
(Abdo et al. 2010a). This will render the Fermi-LAT detectability
marginal even for the four most gamma-ray luminous AXPs 1E
1547.0-5408, XTE J1810-197, 1E 1048.1-5937, and 4U 0142+61. The 17 months exposure
improves the Fermi-LAT sensitivity curve in Figure 1 and 2
quantitatively. Future deeper Fermi-LAT observations are required in
order to make clear this issue.


In addition to the outer gap model, there are also other
high-energy gamma-ray emission mechanisms. For ordinary gamma-ray pulsars, the high-energy gamma-ray
radiation should come from the outer magnetosphere (Abdo et al.
2010a). Outer gap (Cheng 2009), slot gap (Harding 2009), and annular
gap (Qiao et al. 2007), etc, are possible candidates. 
Calculations in other high-energy emission models are called for.

In this paper, we mainly concern about the high-energy gamma-ray properties of AXPs and SGRs.
Tong et al. (2010b) discussed the multiwave properties of 4U 0142+61 as an example.
The demerit of fallback disk model for AXPs and SGRs is that it can not account for the
bursts easily (e.g., Trumper et al. 2010). However, bursts (especially giant flares) in the accretion model are
not absolutely impossible (see discussions of Rothschild et al. 2002; Xu et al. 2006;
Ertan et al. 2007).

As first pointed out by Tong et al. (2010b), the non-detection in Fermi 
observations of all AXPs and SGRs provides challenges to the magnetar model. 
AXPs and SGRs are high-energy gamma-ray emitters in the magentar model because they have strong surface dipole field. 
The strong surface dipole field is the consequence of magnetic dipole braking (e.g., Kouveliotou et al. 1998). 
Since AXPs and SGRs are assumed to be magnetic field powered in the magnetar model, 
it is possible that they have different braking mechanisms to that of rotation powered pulsars 
(e.g., wind braking, Harding et al. 1999). Assuming wind braking for all AXPs and SGRs, the 
corresponding surface dipole field is in the range of normal radio pulsars 
(Harding et al. 1999, they only calculated the case of SGR 1806-20).
This may explain the non-detection of Fermi-LAT observations of all AXPs and SGRs,
at the expenses of dropping the commonly referred magnetic dipole braking assumption and the consequent strong surface dipole field.
The recently discovered low magnetic field SGR (SGR 0418+5729 with $B_{\mathrm{dipole}}<7.5\times 10^{12}\, \mathrm{G}$, Rea et al. 2010)
is consistent with the above analysis. Detailed calculations will be presented in a separate paper.

In the future, if Fermi-LAT can detect high-energy gamma-ray
emissions from one AXP or SGR, it will be strong evidence for
magnetar dipole field ($B_{\mathrm{dipole}}\sim 10^{14}-10^{15}\,
\mathrm{G}$) for this source. This will also open another window for
measuring the effect of strong magnetic fields, i.e., through
unipolar induction effect. This method is independent of timing
measurement, which may be magnetic dipole braking or disk braking.
On the other hand, if still no significant detection is reported in
Fermi-LAT deep observations, it will provide severe challenges to
the magnetar model. From figure 1 and 2, we see that for
many AXPs and SGRs, their theoretical spectra are two or three times
lower than the Fermi-LAT one year sensitivity curve. Four (nine)
years exposure time will make the sensitivity curve two (three)
times lower. Therefore, we expect future four to nine years exposure
time will make clear this issue.

\section{Conclusions}

In this paper, we calculate the application of self-consistent outer
gaps (Zhang \& Cheng 1997; Takata et al. 2010) to the case of
magnetars and compare the results with Fermi-LAT observations of all
known AXPs and SGRs (Abdo et al. 2010b). Our calculations show that
most AXPs and SGRs will emit high-energy gamma-rays and the gap
closure mechanism is dominated by $\gamma-\gamma$ pair production
process, if they are really magnetars. For the most gamma-ray luminous
AXPs 1E 1547.0-5408, XTE J1810-197, 1E 1048.1-5937, and 4U 0142+61, their SEDs are
above the Fermi-LAT sensitivity curve and should have been detected
by Fermi-LAT. The observational upper limits of 4U 0142+61 are below
the theoretical SEDs for large inclination angles. Therefore, there
is conflict between outer gap model (Zhang \& Cheng 1997) in the
case of magnetars and Fermi-LAT observations.

It is possible that AXPs and SGRs are wind braking, i.e., magnetars without a strong surface dipole field (Harding et al. 1999).
It can not be excluded that AXPs and SGRs are fallback disk
systems (Alpar 2001; Chatterjee et al. 2000; Xu et al. 2006). 
Considering the uncertainties in the outer gap modeling (e.g., the solid angle),
future deeper Fermi-LAT observations are required. It will help us
make clear whether AXPs and SGRs are magnetars or fallback disk systems.

\section*{Acknowledgments}
This work is supported by the National Natural Science Foundation
of China (Grant Nos. 10935001, 10973002), the National Basic Research Program of China
(Grant No. 2009CB824800), and the John Templeton Foundation.


\begin{thebibliography}{99}

\bibitem{Abdoa (2010)}
Abdo, A. A., et al. 2010a, ApJS, 187, 460

\bibitem{Abdob (2010)}
Abdo, A. A., et al. 2010b, ApJ, 725, L73

\bibitem{Alpar (2001)}
Alpar, M. A. 2001, ApJ, 554, 1245

\bibitem{Atwood (2009)}
Atwood, W. B., et al. 2009, ApJ, 697, 1071

\bibitem{Beloborodov (2007)}
Beloborodov, A. M., \& Thompson, C. 2007, ApJ, 657, 967

\bibitem{Chatterjee (2000)}
Chatterjee, P., Hernquist, L., \& Narayan, R. 2000, ApJ, 534, 373

\bibitem{Cheng (2009)}
Cheng, K. S. 2009, ApSS, 357, 481

\bibitem{Cheng (1986)}
Cheng, K. S., Ho, C., \& Ruderman, M. 1986, ApJ, 300, 500

\bibitem{Cheng (2001)}
Cheng, K. S., \& Zhang, L. 2001, ApJ, 562, 918

\bibitem{Domerest (2010)}
Demorest, P. B., Pennucci, T., Ranson, S. M., Roberts, M. S. E., \& Hessels, J. W. T. 2010, Nature, 467, 1081

\bibitem{Ertan2004}
Ertan, U., \& Cheng, K. S. 2004, ApJ, 605, 840

\bibitem{Ertan (2007)}
Ertan, U., Erkut, M. H., Eksi, K. Y., \& Alpar, M. A. 2007, ApJ, 657, 441

\bibitem{Frank}
Frank, J., King, A., \& Raine, D. 2003, Accretion power in
astrophysics, Cambridge University Press, Cambridge

\bibitem{Harding 1999}
Harding , A. K., Contopoulos, I., \& Kazanas, D. 1999, ApJ, 525, L125

\bibitem{Harding (2009)}
Harding, A. K. 2009, ApSS, 357, 521

\bibitem{Hirotani (2006)}
Hirotani, K. 2006, ApJ, 652, 1475

\bibitem{Hurley (2009)}
Hurley, K. 2009, ApSS, 357, 575

\bibitem{Kouveliotou}
Kouveliotou, C., et al. 1998, Nature, 393, 235

\bibitem{Lyutikov (2006)}
Lyutikov, M, \& Gavriil, F. P. 2006, MNRAS, 368, 690

\bibitem{Medin}
Medin, Z., \& Lai, D. 2010, MNRAS, 406, 1379

\bibitem{Mereghetti (2008)}
Mereghetti, S. 2008, A\&ARv, 15, 225

\bibitem{Qiao (2007)}
Qiao, G. J., Lee, K. J., Zhang, B., Wang, H. G., \& Xu, R. X. 2007, ChJAA, 7, 496

\bibitem{Rea}
Rea, N., et al. 2010, Science, 330, 944

\bibitem{Rothschild (2002)}
Rothschild, R. E., Lingenfelter, R. E., \& Marsden, D. 2002, 
In: Neutron Stars in Supernova Remnants, ASP Conference Series, 271, 257 (astro-ph/0112121)

\bibitem{Mus (2010)}
Sasmaz Mus, S., \& Gogus, E. 2010, ApJ, 723, 100

\bibitem{Shapiro (1983)}
Shapiro, S. L., \& Teukolsky S. A. 1983, Block holes, white dwarfs, and nuetron stars, John Wiley \& Sons, New York

\bibitem{Takata (2010)}
Takata, J., Wang, Y., \& Cheng, K. S. 2010, ApJ, 715, 1318

\bibitem{TD (1995)}
Thompson, C., \& Duncan, R. C. 1995, MNRAS, 275, 255

\bibitem{TD (1996)}
Thompson, C., \& Duncan, R. C. 1996, ApJ, 473, 322

\bibitem{TLK (2002)}
Thompson, C., Lyutikov, M., \& Kulkarni, S. R. 2002, ApJ, 574, 332

\bibitem{Tonga (2010)}
Tong, H., Xu, R. X., Peng, Q. H., \& Song, L. M. 2010a, RAA, 10, 553

\bibitem{Tongb (2010)}
Tong, H., Song, L. M., \& Xu, R. X. 2010b, ApJ, 725, L196

\bibitem{Trumper (2010)}
Trumper, J. E., Zezas, A., Ertan, U., \& Kylafis, N. D. 2010, A\&A,
518, 46

\bibitem{Wang2006}
Wang, Z. X., Chakrabarty, D., \& Kaplan, D. L. 2006, Nature, 440,
772

\bibitem{Wang (2006)}
Wang, Y., Takata, J., \& Cheng, K. S. 2010, ApJ, 720, 178

\bibitem{Xu (2006)}
Xu, R. X., Tao, D. J., \& Yang, Y. 2006, MNRAS, 373, 85

\bibitem{Xu (2007)}
Xu, R. X. 2007, Advances in space research, 40, 1453

\bibitem{ZhangBing}
Zhang, B. 2003, In: Stellar astrophysics - a tribute to Helmut A. Abt, Astrophysics and Space Science Library, 298, 27 
(astro-ph/0212016)

\bibitem{Zhang (1997)}
Zhang, L., \& Cheng, K. S. 1997, ApJ, 487, 370

\bibitem{Zhang (2002)}
Zhang, L., \& Cheng, K. S. 2002, ApJ, 579, 716

\end{thebibliography}
\end{document}